\documentclass[epj]{svjour}

\usepackage{graphics}
\usepackage{epsfig}

\newcommand{\be}{\begin{equation}}
\newcommand{\ee}{\end{equation}}
\newcommand{\bea}{\begin{eqnarray}}
\newcommand{\eea}{\end{eqnarray}}
\newcommand{\bfig}{\begin{figure}}
\newcommand{\efig}{\end{figure}}
\newcommand{\bc}{\begin{center}}
\newcommand{\ec}{\end{center}}
\newcommand{\btab}{\begin{tabular}}
\newcommand{\etab}{\end{tabular}}

\newcommand{\szz}{\sigma_{zz}}
\newcommand{\sxx}{\sigma_{xx}}

\newcommand{\ra}{\right >}
\newcommand{\la}{\left <}

\newcommand{\dy}{\raisebox{1.6ex}{\rotatebox{180}{\textsf{Y}}}}
\newcommand{\uy}{\textsf{Y}}

\begin{document}

\title{Force chain splitting in granular materials:\\
a mechanism for large scale pseudo-elastic behaviour}

\author{
Jean-Philippe Bouchaud\inst{1}
\and
Philippe Claudin\inst{2}\thanks{\emph{Present address:}
Laboratoire des Milieux D\'esordonn\'es et H\'et\'erog\`enes,
4 place Jussieu - case 86, 75252 Paris Cedex 05, France.}
\and
Dov Levine\inst{2}
\and
Matthias Otto\inst{1}
} 

\institute{
Service de Physique de l'Etat Condens\'e,
CEA-Saclay, Orme des Merisiers,
91191 Gif-sur-Yvette Cedex, France
\and 
Technion - Israel Institute of Technology,
Physics department, Haifa 32000, Israel.
}

\date{\today}

\abstract{
We investigate both numerically and analytically the effect of strong
disorder on the large scale properties of the hyperbolic equations
for stresses proposed in \protect\cite{bcc,wcc}. The physical mechanism that
we model is the local splitting of the force chains (the characteristics
of the hyperbolic equation) by packing defects. In analogy with the theory 
of light diffusion in a turbid medium, we propose a Boltzmann-like equation 
to describe these processes. We show that, for isotropic packings, the 
resulting large scale effective equations for the stresses have exactly the
same structure as those of an elastic body, despite the fact that no
displacement field needs to be introduced at all. Correspondingly, the
response function evolves from a two peak structure at short scales to a
broad hump at large scales. We find, however, that the Poisson ratio is
anomalously large and incompatible with classical elasticity theory that
requires the reference state to be thermodynamically stable.
} 

\authorrunning{J.-P. Bouchaud \emph{et al.}}

\titlerunning{Force chain splitting in granular materials:
a mechanism for a pseudo-elastic behaviour}

\maketitle

\section{Introduction}
The mechanics of assemblies of hard, cohesionless grains has been the subject
of  rather controversial studies in the past few years
\cite{deG,Savage,us,Proc}. A well developed approach, in particular for
engineering applications, is  based on the assumption that granular materials
behave, on large length scales, as elastic (or elastoplastic) bodies. The
microscopic justification of this assumption  for undeformable cohesionless
particles is however far from obvious. The conventional theory of elastic
bodies starts from the identification of a reference state from which
deformations  are defined; the free-energy of the deformed body is then
expanded in powers of the strain  tensor \cite{LL}. The definition of a
smooth deformation field and of its energy are both problematic in the
context of hard grains \cite{Proc}. Recent numerical simulations furthermore
suggest that a limiting curve relating stresses and deformations for large
enough volumes might not even exist \cite{Roux}. 

The absence of any obvious deformation field from which the stress tensor may
be constructed has motivated an alternative, `stress-only' approach
\cite{bcc,wcc,us,these}. The basic tenet of these theories is that in
equilibrium, some (history dependent) relations between the components of the
stress tensor are established. A well known relation of this type arises from
the assumption that the material is everywhere on the verge of plastic failure, leading to a Mohr-Coulomb relation between the stress components
\cite{Nedderman}. A simpler relation,  based both on symmetry arguments and
on the consideration of simple rules for the transfer of stresses between 
adjacent grains, is a local proportionality between the diagonal components
of the stress tensor \cite{bcc}. For example, in two dimensions, this
relation reads $\szz=c_0^2 \, \sxx$, where $c_0$ is a constant. This equation is a local version of the
hypothesis made by Janssen in his famous theory for stresses  in silos
\cite{Nedderman,these,Vanelsilo}. The consequence of this apparently innocuous
assumption is that stresses obey an {\it hyperbolic} equation, as compared to
the elliptic equations encountered in elasticity theory. This means that
stresses `propagate' along lines: the characteristics of this hyperbolic
equation were argued to be the mathematical transcription of  the {\it force
chains} that are well known to exist in granular materials \cite{prl}. More
elaborate versions of the above closure relation between the components of the
stress tensor have been investigated in \cite{wcc,these}, and have been shown
to reproduce both the pressure `dip' underneath the apex of a sandpile
\cite{Smid,Vaneltas}, or the pressure profile inside silos \cite{Vanelsilo}.

A rather clear-cut difference between hyperbolic and elliptic equations lies
in the structure of the propagator, i.e. the response of the stress to a localized force
applied at the top of a layer of granular matter of height $h$. For an
hyperbolic equation, the pressure profile at the bottom of the layer is
localized  around two peaks (in two dimensions) or a circular annulus (in
three  dimensions), centred underneath the point where the force is applied
\cite{bcc}. For an elliptic equation, by contrast, the response function
is a belly shaped curve of width $\propto h$ that reaches its maximum
precisely underneath the point where the force is applied. A similar single
peak structure, but of width $\propto \sqrt{h}$, is also predicted by scalar
models \cite{qmodel}, where  the stress obeys a (parabolic) diffusion
equation. 

This response function has only very recently been carefully measured
experimentally. For well ordered packings, or for frictionless beads, the
two peak structure is rather convincingly observed both in two and three
dimensions \cite{manip2D,chicago}, and has been further  justified
theoretically in \cite{Witten}. However, as soon as the packing is disordered
(for example by considering mixtures of grains of different sizes), the two
peak structure disappears on large length scales and is replaced by a one
peak, elastic like, response function \cite{Clement,manip2D,photo}, with a
width scaling as $h$. This
single hump response function has been also observed in numerical simulations
of disordered packing in \cite{Moreau}. Finally, a purely diffusive response
function (scaling as $\sqrt{h}$) was reported in \cite{JER}, but on a very
special `brick' packing.

The aim of this paper is to investigate both numerically and theoretically
the consequences of strong disorder on the  large scale properties of the
locally hyperbolic equations proposed in  \cite{bcc,wcc}. We propose and 
simulate numerically a simple
model where force chains (characteristics) are  `scattered' in a random way by
defects.  The response function is seen to
evolve, as a function of the height $h$ of the layer, from a two peak
structure at small scales to a one peak elastic like response at large scales.
We then propose a Boltzmann-like formalism to  describe the statistics of this
force chain scattering process. We show that the large scale effective
equations for the stresses become  elliptic, and, rather surprisingly, have
{\it exactly the same structure as the equations  governing the stresses in 
an elastic body}. This is quite striking because no displacement field needs
to be  introduced at all. (This is actually also true of our numerical model
of where  only forces are considered). Interestingly, it is the very existence
of force  chains that allows one to define a local vectorial `order
parameter' (that we call $\vec J$ below) which formally plays the role of
the displacement in elasticity theory. The  importance of force chains for
the mechanics of granular materials was emphasized in \cite{EandO,prl}; the
present work suggests that their presence might be (somewhat paradoxically)
crucial to allow for a large scale elastic like behaviour.

The Poisson ratio $\nu$ that appears in our elastic-like equations depends
solely on  the statistical properties of the force chain splitting process.
Interestingly, this Poisson  ratio violates the bound $\nu \leq 1/2$ (in three
dimensions) that
comes from the fact that the elastic energy of  deformations is a positive
quadratic form of the strain tensor. This clearly shows that  although the
final equations have the same formal structure as those of elasticity theory, 
granular materials do not behave as standard elastic bodies, even for
statistically isotropic packings.

\section{The force splitting model: definition and numerical simulations}

Pictures of photoelastic grains under compression clearly reveal the
existence of  linear force chains which tend to split upon meeting vacancies
or packing defects: see figure \ref{fig1} \cite{Dantu,Radjai}. 
\bfig[t]
\bc
\epsfxsize=\linewidth
\epsfbox{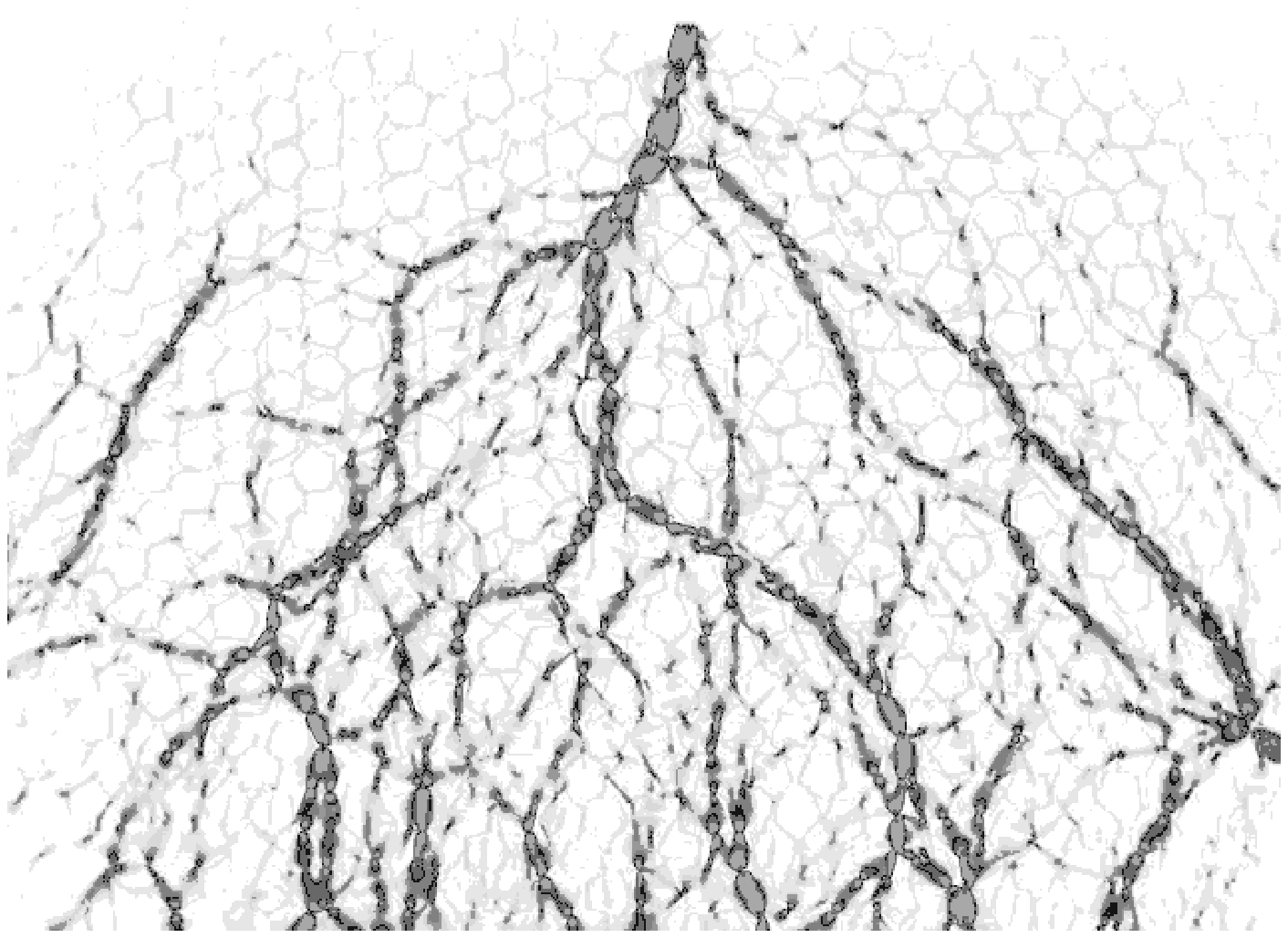}
\caption{Picture of the force chains in a two-dimensional system of grains
subject to a vertical force imposed at the middle of the top surface
\protect\cite{photo}. Such a picture has been obtained with birefringent
grains between perpendicular polarizers.
\label{fig1}}
\ec
\efig
This motivates the following model to compute the response
function $G(x,h)$ to a force imposed at the top of the layer ($x=0$, $z=0$),
and measured at $x$ and $z=h$. For a perfectly ordered packing, we want this
response to be made of two sharp peaks  centered at $x=\pm c_0 h$, where
$c_0=\tan\theta_0$ is the tangent of the angle of the direction along which
forces are transmitted in regular arrangement of grains. For a compact
triangular packing, this angle is $\theta_0=30^o$, which is the value that we
choose  in the following simulations. Therefore we start with two force chains
at angle $\pm\theta_0$ to the vertical -- the initial point $x=z=0$ itself can been
seen as a `defect', see below.

For disordered packing, we randomly distribute non overlapping defects of
radius $a=1$ (this fixes the unit length) with a density $\rho_d$. If one
of the force chains meets a defect, we split it into two new ones at random
angle, which then propagate until another defect (or the boundary) is reached,
an so on. 

More precisely, a chain carrying a force $f$ in the direction $\vec n$ splits
when meeting a defect into two forces $f_1$, $f_2$ in the directions $\vec
n_1$, $\vec n_2$.  The two angles $\theta_1$ and $\theta_2$ (between $\vec
n$ and $\vec n_1$ and $\vec n_2$ respectively) are uniformly chosen between $0$
and some maximum splitting angle $\theta_M$.  The local mechanical equilibrium
imposes that the intensities $f_1$ and $f_2$ are such  that $f \vec n= f_1
\vec n_1+ f_2 \vec n_2$ -- see figure \ref{fig2} (a). Sometimes, two  (or
more) force chains meet at the same defect. In this case, we make them all
merge together -- see figure \ref{fig2} (b). It is important to note that the
positions of the defects are fixed before starting the computation of the
forces. This idea of a frozen disorder is consistent with the experimental
observation that when the local overload is added on the top of the system,
the forces are transmitted along the chains originally created during the
building of the packing. In other words, as long as the applied force is not
too large and compatible with the pre-existing network of force chains, the
geometry of the packing, and in particular the contacts between grains,
remains the same.

\bfig[t]
\bc
\epsfysize=5cm
\epsfbox{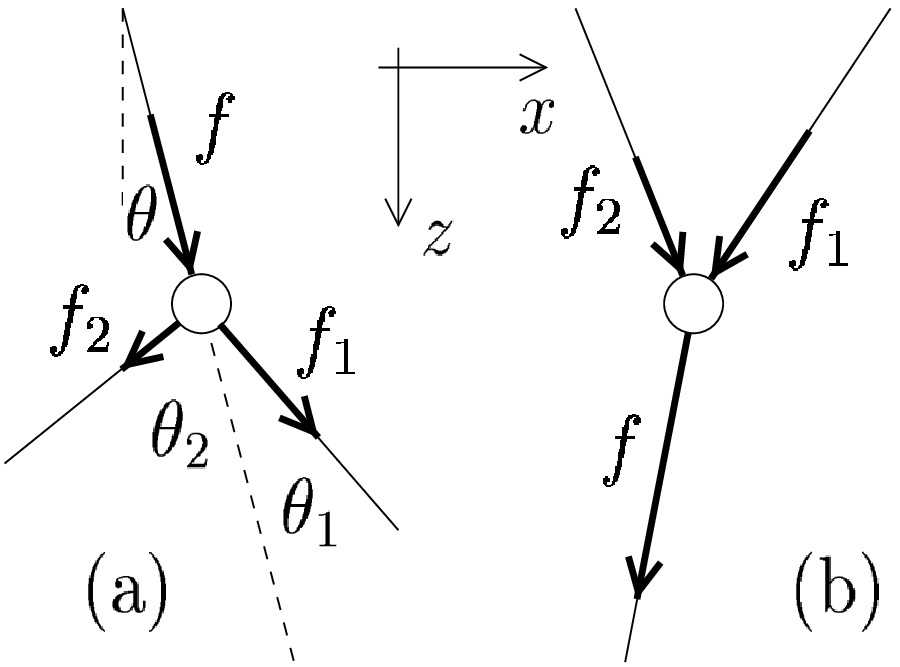}
\caption{The disorder of the packing is modelled by the presence of
`defects' where force chains can split at random angles (a)
(`\protect\dy\ processes') -- or merge (b) (`\protect\uy\ processes'). The
force chains have no width, but the defects have a finite radius $a=1$. At
each defect, forces are balanced: $f \vec n= f_1 \vec n_1+ f_2 \vec n_2$ .
\label{fig2}}
\ec
\efig

The network of forces is built in a hierarchical manner: each defect involved
in this network has one or more `parents' and one or more `children'. The most
common situation is when a defect has one parent and two children. We qualify
such defects to be of `\dy\ type'. Defects of `type \uy' have two (or very 
rarely three or more) parents but one child only. Finally, some defects are
not involved in the network of forces. The type of the defects is not fixed a
priori. At the beginning of the calculation, all defects are `free' and become
progressively of \dy\ type when reached by a force chain. When a force chain
reaches a defect which is already of \dy\ type (or less likely of \uy\ type),
all force chains originally descending from this defect are removed, and it
becomes of \uy\ type. Note that this procedure does not allow cycles, i.e.
defects cannot be their own ancestors -- or descendants \footnote{
Even if cycles were to be allowed, the problem would converge very
quickly -- there is no sense in which allowing cycles would modify,
except infinitesimally, the results obtained.}. We checked that the
final network of force chains is independent of the order in which defects
are chosen at each step. Eventually, all the force chains reach a boundary of
the system, and the calculation stops. All four walls are perfect absorbing
boundaries. 

A further technical difficulty is the presence of very small forces. The deeper
the layer, the larger the number of force chains that carry extremely weak
forces. In order to keep the computation time to reasonable values, the total
number of force chains cannot grow indefinitely. We therefore introduced a
lower cut-off, below which force chains are no longer taken into account. We
checked that for small enough cut-off our results are
independent of this cut-off.

\bfig[t]
\bc
\epsfxsize=\linewidth
\epsfbox{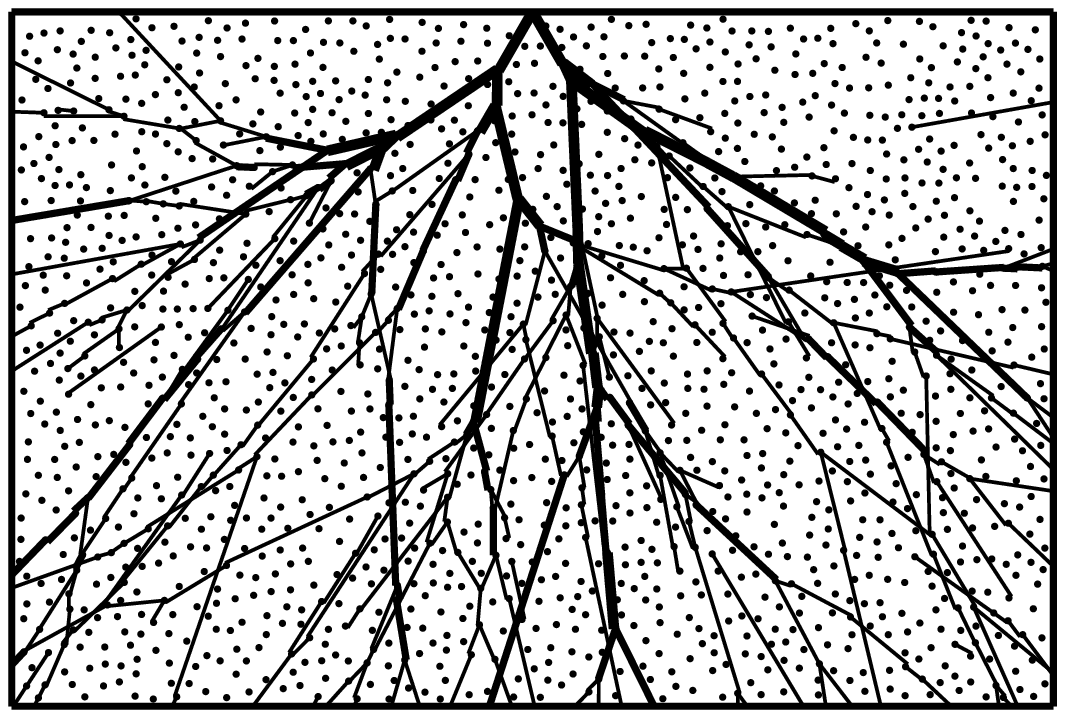}
\caption{Snapshot of the network of force chains obtained within the
\protect\dy\protect\uy\ model. The dots represent the defects of the
granular packing. The lines are bolder when the forces are larger -- very
small forces (less than $0.01$) have not been plotted. In this picture
we set the height to $h=200$, the width to $2w=300$, the density of defects
to $\rho_d=0.028$ and the maximum splitting angle to $\theta_M=30^o$.
\label{fig3}}
\ec
\efig

With these rules, realistic force networks can be created -- see figure
\ref{fig3}. After averaging over many statistically identical samples, one can
obtain stress profiles for different heights $h$.
\bfig[ht!]
\bc
\epsfxsize=\linewidth
\epsfbox{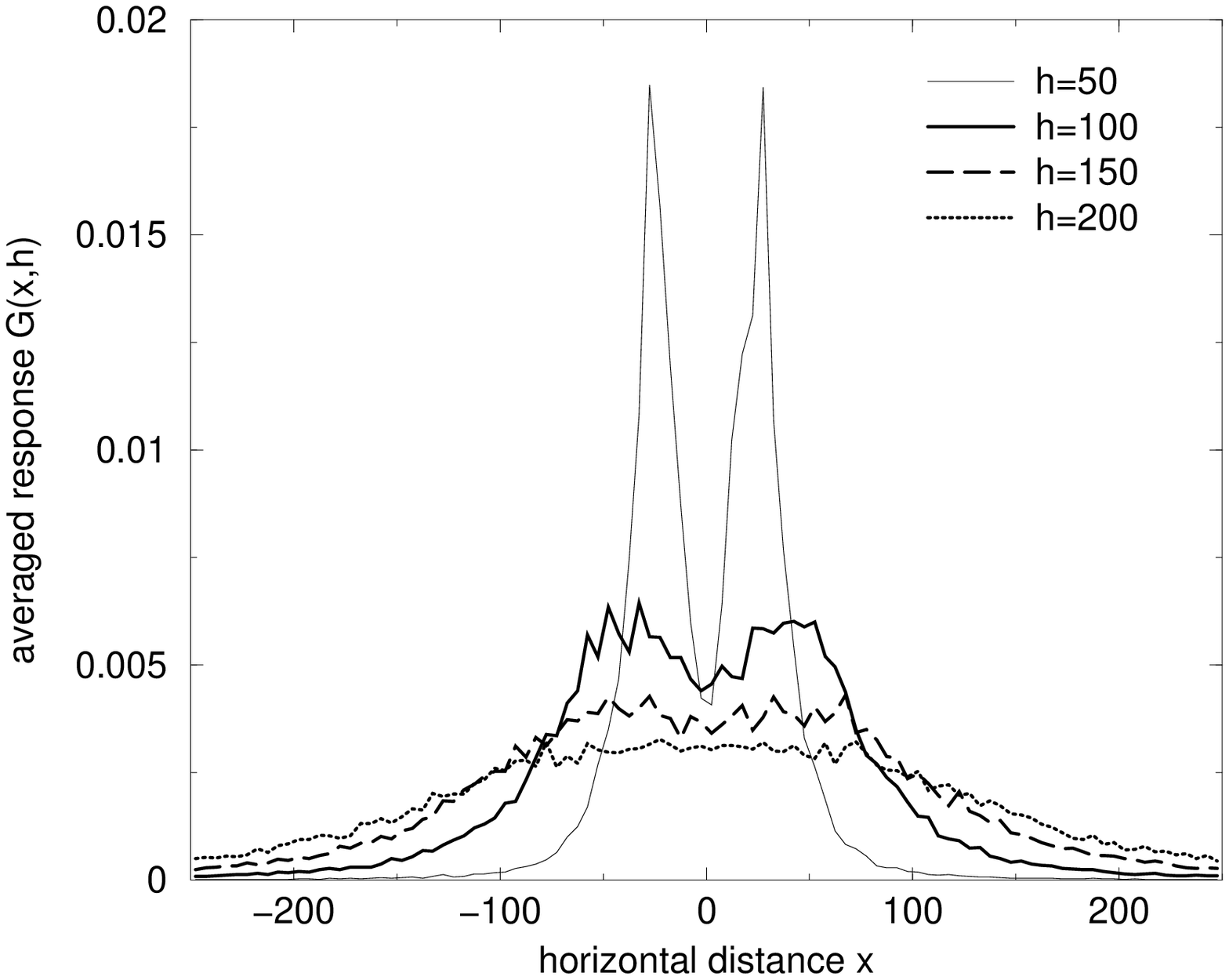}
\caption{Averaged vertical stress response function for different values
of $h$. At small scale, the response function has a double peak shape which is
characteristic of a locally hyperbolic behaviour. For larger heights, these two
peaks merge into a single broad peak, comparable to an elastic-like
response. These curves have been obtained with $2w=500$, $\rho_d=0.028$,
$\theta_M=30^o$, and averaged over $1000$ samples.
\label{fig4}}
\ec
\efig
\bfig[ht!]
\bc
\epsfxsize=\linewidth
\epsfbox{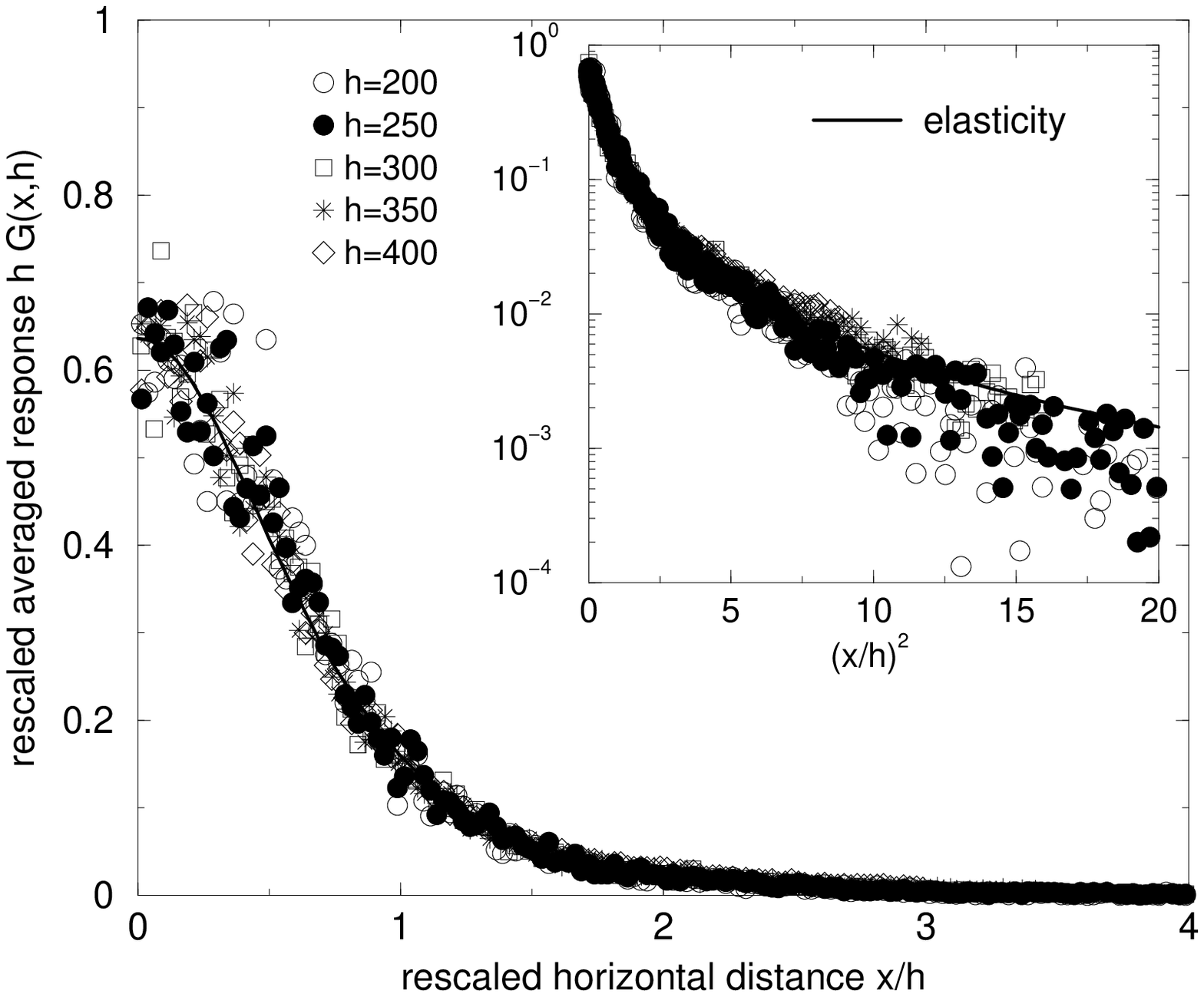}
\caption{For large heights, the width of the response function grows like $h$. 
The inset shows that its shape is not gaussian and actually rather close to a
purely elastic response, shown as the plain line. These curves have been
obtained with $2w=2400$, $\rho_d=0.028$, $\theta_M=30^o$, averaged over $100$
samples, and their integrals normalized to unity.
\label{fig5}}
\ec
\efig
Figure \ref{fig4} shows that, as $h$ increases, the vertical pressure response
profile evolves continuously from two well defined peaks to a single broad one.
It means that the hyperbolic behaviour is progressively erased by multiple
scattering. Besides, the width of the single peak scales like $h$ -- see
figure \ref{fig5}. This scaling is in very good agreement with experiments
\cite{manip2D,Clement,photo} and rules out an analysis both in terms of  the
so-called telegrapher's equation, which was proposed in \cite{Kenkre} to
interpolate between a locally hyperbolic behaviour and a large scale diffusive
 behaviour implied by scalar descriptions \cite{qmodel,JER}. 

More precisely, we show that the shape of this peak is not gaussian: see 
the inset of figure \ref{fig5}. Rather,
it is surprisingly close to the pure elastic response of a semi-infinite
medium, which reads \cite{LL}:
\be
G_e(x,h) = \frac{2}{\pi h} \, \frac{1}{[1 + (x/h)^2]^2},
\label{repelas}
\ee
independently of both the Young modulus and the Poisson ratio. Note that,
because some force chains are deflected upwards and reach the top of the
system,  the integral of the vertical pressure over the bottom wall $z=h$
gives a result slightly larger than $1$  (typically $1.07$ for $h=400$ and
$\theta_M=30^o$). For a proper comparison with the elastic response, we have
renormalized the data so that its integral is unity. However, one should
emphasize that the `backward' propagation of forces is a realistic effect.
The top layer only remains stable as long as these forces do not  exceed the weight
of a single grain. For larger applied forces, one expects the top layer to
become unstable. Let us now turn to an analytical derivation of the numerical
results obtained above.

\section{A Boltzmann equation for force chain splitting}

The role of disorder on hyperbolic equations was previously investigated in
\cite{pre}, where the constant $c_0$ in the equation relating $\szz$
and $\sxx$ was allowed to be randomly fluctuating in space. The study
was restricted to small  fluctuations (i.e. $c_0(\vec r)=
\overline c_0+ \delta c(\vec r)$, with $\delta c \ll \overline c_0$), 
where perturbation theory is valid. In this case, the two peak structure of
the response function is preserved on large length scales, although the peaks
are broadened. An uncontroled extrapolation to strong disorder however 
suggested that
the large scale equations might become elliptic. 

The chain splitting model proposed above allows us to investigate the strongly 
disordered situation. Our strategy is to write a Boltzmann equation for the 
probability density $P(f,\vec n,\vec r)$ of finding an oriented force chain
of intensity $f$ in the  direction $\vec n$ around the point $\vec r$. A very
important point here is that force chains can be oriented in reference to
the boundary conditions. It is indeed a specific
property of hyperbolic equations to require boundary conditions either in the
`past' or in the `future', but not in both  (see the discussion in
\cite{Proc,Witten}). The possibility to orient the chains  has been implicitly
used in the above numerical scheme. 

For simplicity, we shall in the following neglect the chain `merging' 
\uy\ process,
and furthermore that the splitting is symmetric, i.e that $\vec n\cdot\vec
n_1=\vec n\cdot\vec n_2 \equiv \cos \theta$, so that $f_1=f_2=f/2\cos\theta$.
As will be clear from the treatment below, more complicated  cases, with more
than two offsprings or asymmetric scattering do not change the structure  of
the  large scale equations, provided the scattering is isotropic on
average.  (Anisotropic scattering can arise in granular materials with a non
trivial contact texture, and leads  to very interesting generalisations that
will be mentioned in the conclusion). 

Assuming a uniform density of defects, the probability distribution $P(f,\vec n,\vec r)$ obeys the
following general equation:
\bea
\lefteqn{P(f,\vec n,\vec r+\vec n \, dr)=
(1-\frac{dr}{\lambda}) P(f,\vec n,\vec r) \, +} & \nonumber \\
& & 2 \frac{dr}{\lambda} \! \int \!\! d\vec n' \!\! \int \!\! df'
P(f',\vec n',\vec r) \Psi(\vec n',\vec n)
\, \delta \! \left(f-\frac{f'}{2\cos\theta}\right)\!,
\label{Boltzmann}
\eea
where $\lambda$ is equal to the `mean free path' of force chains, and is of
order $1/(\rho_d a^{D-1})$ in dimension $D$. The above equation means
the following: since a chain of grains can only transport  a force parallel to
itself \cite{prl}, the direction of the force $\vec n$  also gives the local 
direction of the chain. Between $\vec r$ and $\vec r+\vec n \, dr$, the chain
can either  carry on undisturbed, or be scattered. The second term on the
right hand side  therefore  gives the probability that a force chain initially
in direction $\vec n'$ is scattered in direction $\vec n$. This occurs with
a probability $\frac{dr}{\lambda} \Psi(\vec n',\vec n)$, where 
$\Psi$ is the scattering cross section, which we will assume to depend 
only on the
scattering angle $\theta$, for example a uniform distribution between 
$-\theta_M$ and $+\theta_M$. The $\delta$-function ensures force conservation 
and the  factor two comes from the counting of the two possible outgoing
force chains.

Let us now multiply Eq. (\ref{Boltzmann}) by $f$ and integrate over $f$. This 
leads to an equation for the local average force per unit volume in the
direction $\vec n$, that we will denote $F(\vec n,\vec r)$. This equation
reads:
\bea
\lefteqn{\lambda \, \vec n \cdot \! \vec \nabla_r F(\vec n,\vec r)=
- F(\vec n,\vec r) \, +} & \nonumber \\
& & \int \!\! d\vec n' \, \frac{F(\vec n',\vec r)}{\vec n \cdot \vec n'} \,
\Psi(\vec n',\vec n) + \frac{\lambda}{a} \, \vec n \cdot \vec F_0(\vec r),
\label{SM}
\eea
where we have added the contribution of an external body force density $\vec
F_0(\vec  r)$, and $a$ is the size of a defect or of a grain. This
equation is the so-called Schwarschild-Milne equation for radiative  transfer,
decribing the evolution of light intensity in a turbid medium \cite{Theo}. We
now introduce some angular averages of $F(\vec n,\vec r)$ that have an
immediate physical  interpretation:
\bea
p(\vec r) & = & a \! \int \!\! d\Omega \, F(\vec n,\vec r) \\
J_\alpha(\vec r) & = & a \int \!\! d\Omega \ n_\alpha \, F(\vec n,\vec r) \\
\sigma_{\alpha \beta}(\vec r) & = & a D \! \int \!\! d\Omega \ n_\alpha  n_\beta
\, F(\vec n,\vec r),
\eea
where $\int \!\! d\Omega$ is the normalized integral over the unit sphere, that
is introduced for a correct interpretation of $\sigma$ (see Eq. (\ref{second}) 
below). As will
be clear from the following, $\vec J$ is the local
average force chain intensity per unit surface, $\sigma$ is the stress tensor.
Since $\vec n^2=1$, one finds that Tr$\sigma=Dp$, and therefore $p$ is the 
isostatic pressure. Note that $\vec J$ is not the average local force, which is always
zero in equilibrium. The fact that $\vec J \neq \vec 0$ comes from the 
possibility of orienting the force chains.

\section{The hydrodynamic limit: back to `elasticity'}

Let us indeed integrate over the unit sphere Eq. (\ref{SM}) after multiplying it by 
different powers of $n_\alpha$. Using the fact that $\Psi(\vec n',\vec n)$
only depends on $\vec n \cdot \vec n'$, a direct  integration leads to:
\be
\lambda \, \vec \nabla \! \cdot \vec J = (a_0-1) p,
\label{first}
\ee
where $a_0$ is called the `albedo' in the context of light scattering
\cite{Theo}, and reads:
\be
a_0 \equiv \int \!\! d\vec n \, 
\frac{\Psi(\vec n',\vec n)}{\vec n \cdot \vec n'} \geq 1.
\ee
A second set of equations can be obtained by multiplying by $n_\alpha$ and 
integrating. Using the fact that
\be
a_1 = \int \!\! d\vec n \, \Psi(\vec n',\vec n) = 1,
\ee
it is easy to show that
\be
\int \!\! d\vec n \,\vec n \, \frac{\Psi(\vec n',\vec n)}{\vec n \cdot \vec n'}
= \vec n'.
\ee
Therefore, the resulting equation is nothing but the usual
mechanical equilibrium relation:
\be
\nabla_\beta \sigma_{\alpha \beta} = F_{0\alpha}.
\label{second}
\ee
This relation reflects the local balance of forces chains.
Now we multiply Eq. (\ref{SM}) by $n_\alpha n_\beta$ and again integrate. A 
priori, this introduces a new three index tensor. In order to close the
equation, we now  make the an assumption that is usually made in the context
of light diffusion, that on large scales the force intensity becomes nearly
isotropic \cite{Theo}. In this case, it is justified to expand $F(\vec n,\vec
r)$ in angular harmonics and to keep only the first  terms
\footnote{Actually, the third term in the expansion, proportional to the 
stress tensor, does not contribute to the following closure equation. Only
the fourth order  term  (in $nnn$) would change the following results.}:
\be
a \, F(\vec n,\vec r)=p(\vec r)+  D\vec n \cdot \vec J(\vec r) + ...
\ee
The calculation requires the following 
identity (note that the repeated indices are not summed here):
\bea
\lefteqn{\Gamma_{\alpha \beta \gamma \delta}
\equiv \int \!\! d\Omega \, n_\alpha n_\beta n_\gamma n_\delta =
K_1 \left[ \delta_{\alpha \beta} \delta_{\gamma
\delta} (1-\delta_{\alpha \gamma}) \, + \right.} & \nonumber \\
& & \left. (1-\delta_{\alpha \beta})(\delta_{\alpha \gamma} 
\delta_{\beta \delta} + \delta_{\alpha \delta} \delta_{\beta \gamma}) \right]
+ K_2 \, \delta_{\alpha \beta}\delta_{\alpha \gamma}\delta_{\alpha \delta},
\eea
where $K_1=1/D(2+D)$ and $K_2=3K_1$. The resulting equation finally leads:
\bea
\lefteqn{\lambda D \Gamma_{\alpha \beta \gamma \delta}
\nabla_\gamma J_\delta = } & \nonumber \\
& & \frac{a_0-a_2}{D-1} p \, \delta_{\alpha \beta} +
\left(\frac{Da_2-a_0}{D-1}-1\right) \frac{\sigma_{\alpha \beta}}{D},
\label{third1}
\eea
where 
\be
a_2= \la \cos^2\theta \ra \equiv \int \!\! d\vec n \, {\Psi(\vec n',\vec n)}
\, {\vec n \cdot \vec n'}.
\ee
Taking the trace of Eq. (\ref{third1}), we find, using the fact that 
the trace of $\sigma$ is equal to $Dp$:
\bea
\lefteqn{\lambda 
D[K_1(D-1)+K_2] \, \vec \nabla \! \cdot \vec J = } & \nonumber \\
& & \left( \frac{a_0-a_2}{D-1}D+
\frac{Da_2-a_0}{D-1}-1 \right) p,
\eea
which, using the values of $K_1$ and $K_2$, can be seen to be identical to the
first equation, Eq. (\ref{first}). The final equation appears therefore as a 
`constitutive' relation between $\sigma_{\alpha \beta}$ and the vector
$\vec J$. We find:
\be
\left.\sigma_{\alpha \beta}\right|_{\alpha \neq \beta}=
\frac{\lambda D^2K_1}{\left(\frac{Da_2-a_0}{D-1}-1\right)} \left(\nabla_\alpha
J_\beta +\nabla_\beta J_\alpha \right),
\ee
and
\bea
\lefteqn{\sigma_{\alpha \alpha} =
\frac{\lambda D}{\left(\frac{Da_2-a_0}{D-1}-1\right)}
\left[ D(K_2-K_1) \, \nabla_\alpha J_\alpha \, + \right.} & \nonumber \\
& & \left. \left(DK_1-\frac{a_0-a_2}{(D-1)(a_0-1)} \right)
\vec \nabla \! \cdot \vec J \right]
\eea
(here $\alpha$ is not summed over.) These equations have exactly the canonical form of elasticity theory, 
provided one identifies the vector $\vec J$ with the local displacement, up to a
 multiplicative factor. This free multiplicative factor means that the
effective Young modulus cannot be uniquely defined, as expected since it is
not a dimensionless quantity.  However, the Poisson ratio $\nu$ can be
defined, and reads:
\be \nu=\frac{1}{D-1}\
\frac{D-1+3a_0-(D+2)a_2}{D+1+a_0-(D+2)a_2}.
\ee
Note that this quantity is purely geometric, and does not involve the
scattering mean free path $\lambda$. Since $a_0 \geq 1$ and $a_2 \leq 1$,
it is easy to show that $\nu \geq 1/(D-1)$ in all dimensions,
therefore violating the thermodynamical bound $\nu \leq 1/(D-1)$ that holds for
conventional elastic bodies. This
means that the stress equations cannot be derived from the minimisation of a
free energy functional. Therefore, granular materials (if describable by
the present theory) should have an anomalously high Poisson ratio. The
Poisson ratio for a uniform distribution of scattering angles is plotted as
a function of the maximum angle $\theta_M$ in figure \ref{fig6}, both for
$D=2$ and $D=3$. In the weak scattering limit ($\theta_M \to 0$), one finds
$\nu =(D+5)/(D+3)(D-1)$ independently of the precise shape of $\Psi$. In the strong
scattering limit where $\theta_M \to \pi/2$, outgoing force  chains can
carry arbitrarily large forces and therefore $a_0 \to \infty$. In this case,
$\nu \to 3/(D-1)$. Such anomalous values of the Poisson ratio in granular
materials were discussed in recent papers \cite{Poisson}, although the
underlying mechanisms are different from the one proposed here.

\bfig[t]
\bc
\epsfxsize=\linewidth
\epsfbox{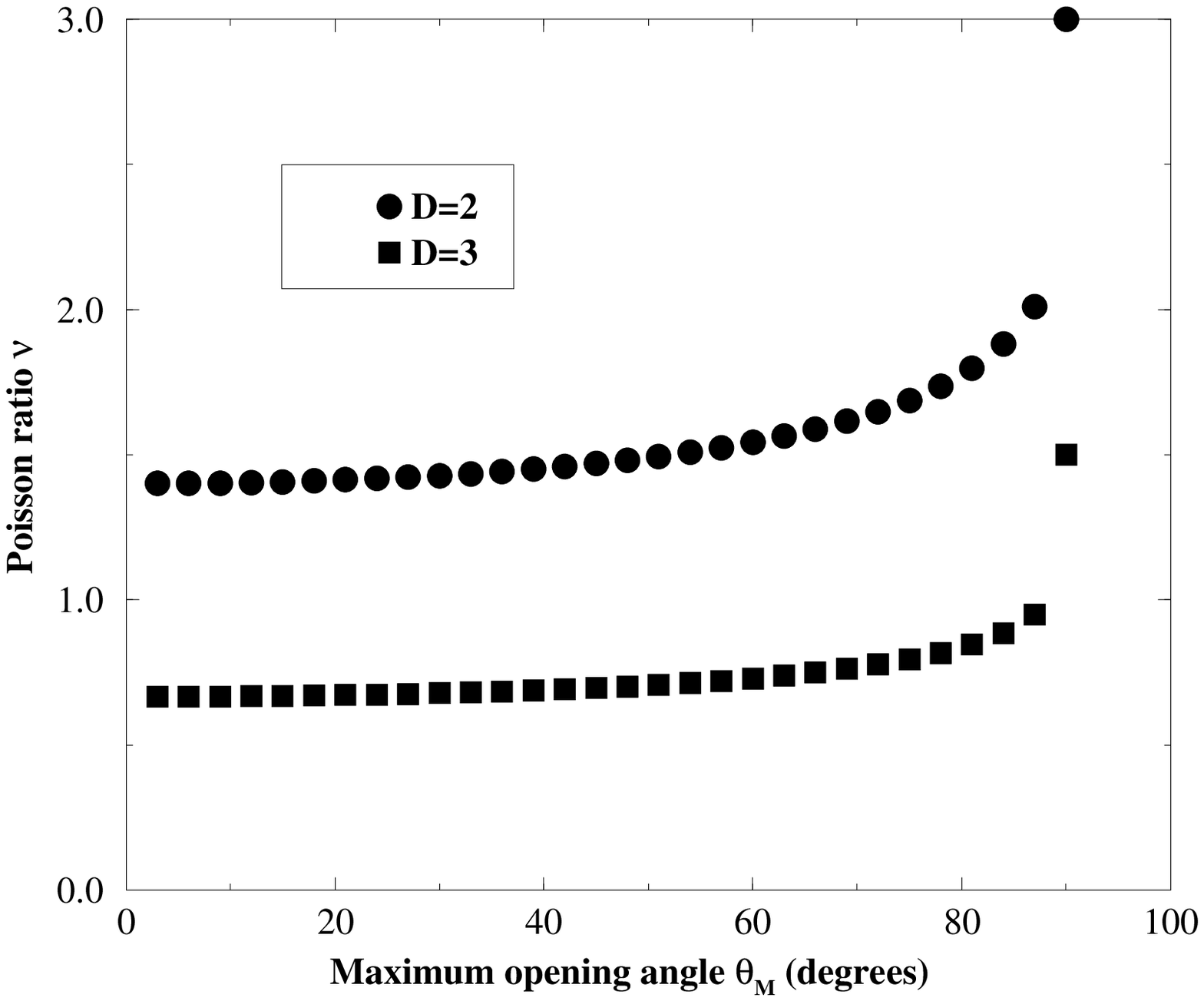}
\caption{Poisson ratio $\nu$ for a uniform distribution of scattering angles,
as a function of the maximum angle $\theta_M$, both for $D=2$ (plain line)
and $D=3$ (dashed line).
\label{fig6}}
\ec
\efig

Since the above equations are formally identical to those of classical 
elasticity, one can show that $\nabla^2 p=0$, and $\nabla^4 \vec J=0$ \cite{LL}.
One can therefore compute the response function $G(\vec r)$ to a 
localized force at $\vec r=0$
in the $z$ direction, which is given in $D=2$ by Eq. (\ref{repelas}), and is,
as figure \ref{fig5} shows, in extremely good agreement with the numerical
simulation of our force chain splitting model. This response function is 
also very close to the one measured experimentally \cite{photo,Clement,manip2D}
or numerically in disordered packings \cite{Moreau}.

\section{Conclusion and extensions}

In this paper, we have shown that in the presence of strong disorder that 
`scatters' the force chains in an assembly of undeformable, cohesionless
grains, the local  hyperbolic character of the stress equations is unstable
at large length scales, where  the equations are akin to those of a
conventional elastic body. This is at variance with  the weak disorder
calculations presented in \cite{pre}, and shows that the unintuitive  two peak 
nature of the response function is perhaps restricted to very ordered packings,
or frictionless particles, for which $\lambda \to \infty$.

Let us stress once more that our result is not trivial because no
displacement field is introduced in the above derivation (nor in our
numerical model).  The basic assumption of our model is  the existence of
local force chains, which have a well defined identity over several grain 
sizes $a$, such that $a \ll \lambda$: this is the condition under which the
above Boltzmann description of the force chain scattering is justified.
Interestingly, these chains that propagate the forces parallel to themselves
are the signature of a locally hyperbolic  equation, and are associated with
its characteristics \cite{prl,these}. 

This work can be extended in several directions. First, and most importantly, 
the extension to non-isotropic scattering should be worked out. This means that 
the scattering function $\Psi(\vec n,\vec n')$ could depend not only on
$\vec n \cdot \vec n'$ but  also on $\vec N \cdot \vec n'$, where $\vec N$ is
a (space dependent) vector that  describes the local texture, i.e. the non-isotropic distribution of contacts between  grains.  In this case,
combinations like $N_\alpha J_\beta$ or $N_\alpha N_\beta$ can appear in the
stress tensor $\sigma_{\alpha \beta}$. The vector $\vec N$ is expected to
encode the construction history of the material, and  should be non trivial,
for example, in a sandpile constructed from a point source. The modification
of the above theory to account for texture effects is needed to understand the
existence of the pressure dip underneath the apex of the pile 
\cite{Smid,Vaneltas}, which cannot be reproduced by an isotropic
elastic theory. In the presence of a long-ranged texture field $\vec N$, the
response function will depart from  that of elasticity theory. Second, the
fluctuations of stresses in granular materials are very important and have
been discussed theoretically and experimentally \cite{qmodel}. The above
theoretical framework should allow to make some progress, by  studying higher 
moments of the local force distribution $P(f,\vec n,\vec r)$, beyond the 
average force $F(\vec n,\vec r)$ studied here. Third, the pseudo-elastic
theory established  above is only expected to hold on length scales such that
$L \gg \lambda$. This does  not necessarily hold for small piles or small
silos, where the hyperbolic  approach seems to give very good results
\cite{Vanelsilo}. It would be interesting to work out in  more details the
crossover regime where $L/\lambda$ is not large compared to one, using for
example the  formalism  developed in \cite{Luck}. Finally, let us note that
the above theory should  describe the large scale properties of random
networks of rigidly jointed struts.

\section*{Acknowledgments}

We wish to thank M. Cates, E. Cl\'ement, J. Duran, 
G. Grest, A. Kamenev, E. Kolb, J.M. Luck,
Y. Nahmias, G. Ovarlez and G. Reydellet for very useful discussions.
P.C. was supported by an Aly Kaufman post-doctoral fellowship. 
D.L. acknowledges support from Israel Science Foundation grant 
211/97 and from U.S. - Israel Binational Science Foundation 
grant 1999235. M.O. is
supported by a DFG research fellowship.

\end{document}